\documentclass[a4paper,fleqn,usenatbib]{mnras}

\usepackage[T1]{fontenc}
\usepackage{aecompl}

\usepackage{graphicx}	
\usepackage{amsmath}	
\usepackage{amssymb}	

\usepackage{xspace}
\usepackage{siunitx}
\usepackage{tikz}
\usepackage{mhchem}

\newcommand{\snia}{SN~Ia}
\newcommand{\snias}{SNe~Ia}

\newcommand{\sngz}{SN~2006gz}
\newcommand{\sndc}{SN~2009dc}
\newcommand{\sndn}{SN~2012dn}
\newcommand{\stella}{\textsc{Stella}\xspace}
\newcommand{\stellastat}{\textsc{Stella-Stat}\xspace}
\newcommand{\artis}{\textsc{Artis}\xspace}
\newcommand{\tardis}{\textsc{Tardis}\xspace}
\newcommand{\mcrh}{\textsc{Mcrh}\xspace}
\newcommand{\sndctail}{\texttt{09dc-tail}\xspace}
\newcommand{\sch}{Super-Chandrasekhar}
\newcommand{\modnocsm}{\texttt{m100101\_nocsm}\xspace}
\newcommand{\modfid}{\texttt{m100101\_csm\_solar}\xspace}
\newcommand{\modztwo}{\texttt{m100101\_csm\_fZ2}\xspace}
\newcommand{\modzfive}{\texttt{m100101\_csm\_fZ5}\xspace}
\newcommand{\modztwenty}{\texttt{m100101\_csm\_fZ20}\xspace}
\newcommand{\modzhundred}{\texttt{m100101\_csm\_fZ100}\xspace}
\newcommand{\modmassivecsm}{\texttt{m100101\_bigcsm\_solar}\xspace}
\newcommand{\modfidearly}{\texttt{m100101\_csm\_solar\_1000}\xspace}
\newcommand{\modfidearlier}{\texttt{m100101\_csm\_solar\_100}\xspace}
\newcommand{\modfidesmall}{\texttt{m100101\_csm\_solar\_100\_R5}\xspace}
\newcommand{\modfidesmaller}{\texttt{m100101\_csm\_solar\_100\_R2}\xspace}
\newcommand{\modfidesmallest}{\texttt{m100101\_csm\_solar\_100\_R1}\xspace}

\newcommand{\parnumin}{\ensuremath{\SI[retain-unity-mantissa=false]{6.10e+13}{Hz}}}
\newcommand{\parnumax}{\ensuremath{\SI[retain-unity-mantissa=false]{2.94e+18}{Hz}}}
\newcommand{\parlammin}{\ensuremath{\SI[retain-unity-mantissa=false]{1.02e+00}{\AA}}}
\newcommand{\parlammax}{\ensuremath{\SI[retain-unity-mantissa=false]{4.91e+04}{\AA}}}

\graphicspath{{./}}

\title[SNe Ia within dense C/O-rich envelopes]{Type Ia supernovae within dense carbon-oxygen rich envelopes: a model for `\sch{}' explosions?}

\author[Noebauer~et~al.]{U.~M.~Noebauer,$^1$\thanks{unoebauer@mpa-garching.mpg.de}
S.~Taubenberger,$^{1,2}$
S.~Blinnikov,$^{3,4,5}$
E.~Sorokina,$^{3,4,6}$
\newauthor
and W.~Hillebrandt$^1$\\
$^1$Max-Planck-Institut f\"ur Astrophysik, Karl-Schwarzschild-Str.~1, D-85748 Garching, Germany\\
$^2$European Southern Observatory, Karl-Schwarzschild-Str. 2, D-85748 Garching, Germany\\
$^3$Institute for Theoretical and Experimental Physics (ITEP), 117218 Moscow, Russia\\
$^4$Kavli Institute for the Physics and Mathematics of the Universe (WPI), The University of Tokyo, Kashiwa, Chiba 277-8583, Japan\\
$^5$All-Russia Research Institute of Automatics (VNIIA), 127005 Moscow, Russia\\
$^6$Sternberg Astronomical Insitute, M.V.Lomonosov Moscow State University,
119991 Moscow, Russia
}

\date{Accepted XXX. Received YYY; in original form ZZZ}

\pubyear{2016}

\begin{document}
\label{firstpage}
\pagerange{\pageref{firstpage}--\pageref{lastpage}}
\maketitle

\begin{abstract}
  We investigate the consequences of fairly normal Type Ia supernovae being
  embedded in compact and dense envelopes of carbon and oxygen rich
  circumstellar material by means of detailed radiation hydrodynamic
  simulations. Our main focus rests on exploring the effects of the interaction
  between ejecta and circumstellar material on the ejecta evolution and the
  broad-band light curve. In our calculations, we find that a strong reverse
  shock efficiently decelerates and compresses the ejecta material. This leads
  to a significant broadening of the optical light curve, a longer rise to
  maximum and a slower decline in the tail phase. During the interaction,
  substantial radiative energy is generated, which mostly emerges in the
  extreme ultraviolet and X-ray regime. Only if reprocessing due to
  radiation--matter interactions is very efficient, a significant boost in the
  optical light curve is observed. We discuss these findings in particular in
  the context of the super-luminous event \sndc{}. As our calculations are able
  to reproduce a number of its peculiar properties, we conclude that the
  flavour of the interaction scenario investigated in this work constitutes a
  promising candidate to explain such `\sch{}' supernovae.
\end{abstract}




\section{Introduction}

The presence of circumstellar material (CSM) is playing an ever more important
role for the understanding of supernova (SN) evolution. For example, the
interplay between CSM and the SN ejecta can be crucial for shaping some of the
defining properties for particular SN classes. Prominent examples in this
context are Type IIn supernovae (SNe IIn).  Here, the eponymous narrow line
features \citep{Schlegel1990} are ascribed to the presence of a dense CSM
envelope, in which hydrogen recombines \citep[e.g.][]{Chugai2004}. The presence
of substantial amounts of CSM is also often invoked as a possible explanation
for the intense luminosity of some of the most powerful SN events
\citep[e.g.][]{Ofek2007,Chevalier2011}, which are typically referred to as
super-luminous SNe. In this scenario, the vast kinetic energy pool of the
ejecta may be tapped through the shock heating processes in the ejecta--CSM
interaction, converted partially to thermal and radiation energy and thus power
the intense light output of such systems (see, for example, the ejecta--CSM
interaction calculation for the super-luminous SN PTF12dam by
\citealt{Baklanov2015}). Note, however, that in the context of super-luminous
SNe, also other models, such as the pair-instability mechanism
\citep[e.g.][]{Barkat1967,Gal-Yam2009} or the magnetar-powered scenario
\citep[e.g.][]{Kasen2010a, Nicholl2013}, are heavily discussed.  The increasing
relevance of ejecta--CSM interaction is also owed to the success of modern
survey programmes in catching SNe at ever earlier phases
\citep[e.g.][]{Gal-Yam2014}. During these epochs, right after the explosion,
the observables probe the immediate vicinity of the explosion site.  Any
interaction with CSM at these times imprints characteristic features onto
spectra and the early light curves (see, for example, systematic exploration in
the \snias{} context by \citealt{Piro2016}). Interpreting these, gives insights
into the mass-loss history of the progenitor system and thus into the exploding
object.

In recent years, an increasing number of observations has revealed CSM interaction
signatures in Type Ia supernovae (\snias{}) as well, which are associated with the
complete thermonuclear incineration of a carbon-oxygen white dwarf (WD). A
prominent examples was PTF11kx \citep{Dilday2012}, which exhibits multiple CSM
shells in its immediate environment, but many more strongly interacting
\snias{} have been identified \citep[see, for example, census
by][]{Silverman2013}. In addition to the direct observational evidence, the
presence of CSM is invoked as a potential explanation for a class of \snias{},
which often exhibit extraordinary luminosities. These events, commonly dubbed
`\sch{}' explosions \citep{Howell2006}, of which \sndc{}
\citep{Yamanaka2009,Silverman2011,Taubenberger2011} is the prototype, elude an
explanation within the standard Chandrasekhar-mass explosion paradigm.

In light of the relevance of the interplay between \snia{} ejecta and its
circumstellar environment, we perform detailed radiation hydrodynamical
calculations of interacting \snias{} in this work. Conceptually similar
explorations have been performed by \citet{Khokhlov1993, Nomoto2005, Fryer2010,
Blinnikov2010}. However, we focus here on a specific realisation of the
interaction scenario, which draws inspiration from previous investigations of
the ejecta--CSM interplay in the context of super-luminous \snias{}
\citep{Taubenberger2013}. In particular, we consider fairly normal \snias{}
occurring within a dense carbon and oxygen rich envelope and examine the
consequences of the ensuing ejecta--CSM interaction for the overall evolution
of the system and its energy output. Apart from determining the generic
evolution of these interaction models, an important aspect of this work lies in
exploring whether this scenario provides a plausible explanation for
\sch{} \snias{}. 

We begin this study by briefly reviewing some key aspects of super-luminous
\snias{} in Section \ref{sec:sn09dclikes}. This is followed by a detailed
overview of the investigated models and the used numerical tool in Section
\ref{sec:model_numerics}. The results of our simulations, which are presented
and compared with observations of super-luminous \snias{} in Section
\ref{sec:results}, will be discussed in detail in Section \ref{sec:discussion}. 

\section{Super-Chandrasekhar supernovae}
\label{sec:sn09dclikes}

\subsection{General Properties}

We briefly review some of the important characteristics of a sub-class of
thermonuclear supernovae of which we consider \sndc{} the prototype. All
objects in this sub-class share some distinct features which set them apart
from `normal' Type Ia explosions, used in cosmological studies.  However,
within the members of this group a sizeable spread in their properties
persists, leading to quite a non-homogeneous class, which we refer to as
`\sch{}' objects.

Considering the photometric appearance, the most striking feature of the \sch{}
objects is their high to outstanding peak luminosity, surpassing the intrinsic
maximum brightness of normal \snias{} by factors up to two
\citep{Howell2006,Scalzo2010,Yuan2010}. In addition to the high luminosity, the
\sch{} objects show a broad and slowly declining light curve with a typical
$\Delta m_{15}(B) \sim 0.6 - 1$ \citep{Scalzo2010,Taubenberger2011}. The light
curve rise to its peak seems to progress slowly, in particular in the case of
\sndc{}, for which a very early detection has determined a lower rise time
limit, $t_{\mathrm{rise}} \ge \SI{23}{d}$ \citep{Silverman2011}. After maximum,
the \sch{} objects exhibit a generically slower decline during the tail phase
than normal \snias{} \citep[e.g.][]{Taubenberger2013}. However, at least in a
number of objects, \sngz{} \citep{Maeda2009}, \sndc{} \citep{Silverman2011,
Taubenberger2011} and possibly \sndn{} \citep{Chakradhari2014}, a prominent
break in the late light curve evolution has been observed, which is often
attributed to the onset of dust formation
\citep{Maeda2009,Taubenberger2011,Nozawa2011,Taubenberger2013}. For \sndc{},
this occurred around $t \sim \SI{200}{d}$ \citep{Taubenberger2011}.

The group of \sch{} objects not only differs from normal \snias{} in terms of
its photometric behaviour but it also shows a number of characteristic spectral
peculiarities. Typically, the spectral energy distribution (SED) of these
objects peaks at bluer wavelengths than expected for normal \snias{} during
early phases \citep{Brown2014}. Related to this, the absorption troughs of
prominent P-Cygni lines, in particular of Ca~II and Si~II, are not very deep
which may be interpreted as an indication for an underlying continuum
contribution \citep{Hachinger2012}.  Contrary to normal \snias{} for which
carbon lines are typically weak and only observed during the early phases, such
lines appear much stronger and persist also until later times in the \sch{}
class \citep[e.g.][]{Taubenberger2011,Chakradhari2014}. Finally, the line
velocities, of important \snia{} lines, as measured by the blue shift of the location of maximum
absorption, exhibit a peculiar behaviour: in
\sndc{}, these velocities are consistently lower than in normal \snias{}
\citep{Yamanaka2009,Silverman2011,Taubenberger2011}. However, a complementary
behaviour may also be observed in this class: the less luminous members, such
as SN~2006gz \citep{Hicken2007} and SN~2012dn \citep{Chakradhari2014}, show
typically line velocities which are much closer to those observed in normal
\snias{}.

\subsection{Proposed Models}

Many of the peculiar properties listed above challenge the canonical
theoretical model for \snias{}. In particular, the high intrinsic luminosity of
the \sch{} objects stands in contrast to the standard interpretation, in
which the light output from \snias{} is solely attributed to the energy
released in radioactive decay reactions. A number of theoretical scenarios have
been discussed in the literature to address this discrepancy. In general, these
models may be divided into two groups: one attempts to reconcile these
events with the standard radioactivity picture by exploring
mechanisms to increase the stability limit of white dwarfs. In contrast to
that, approaches of the second group search for additional energy sources,
which may contribute to the light output in these super-luminous events. 

In general, a nickel mass of $M_{\mathrm{Ni}} \approx 1.8 \, \mathrm{M}_{\sun}$
is needed to explain the peak luminosity of \sndc{} solely with the energy
release from radioactivity \citep{Yamanaka2009, Silverman2011,
Taubenberger2011}. Synthesising that amount of radioactive material during the
thermonuclear burning phase would in turn require an even higher total mass of
the progenitor white dwarf. \citet{Taubenberger2011} estimate an ejecta mass of
the order of $M_{\mathrm{ej}} \approx 2.8\,\mathrm{M}_{\sun}$, which clashes
with the classical Chandrasekhar mass, the stability limit of a degenerate
electron gas. One possibility to overcome this limit is to invoke rotation. A
rapidly rotating WD may reach masses up to $\sim 2.1\, \mathrm{M}_{\sun}$
\citep{Yoon2005}.  Detailed numerical simulations
\citep{Pfannes2010,Pfannes2010a} have demonstrated that pure detonations may
efficiently incinerate such heavily rotating white dwarfs, producing large
amounts of radioactive nickel. However, the intense thermonuclear burning leads
to high kinetic energies in the ejecta and iron group and intermediate mass
elements being located at rather high velocities, contrary to what is observed.
Moreover, even in the very extreme case investigated by \citep{Kamiya2012}, in which
rotating WD models with $M = 2.8\,\mathrm{M}_{\sun}$ have been constructed, the
light curve peak of \sndc{} could only be reproduced after neglecting host galaxy
extinction. In a complementary scenario \citep[suggested for example
by][]{Howell2006,Hicken2007}, the classical stability limit is overcome by
considering the merger of two heavy (possibly rotating) WDs. The total mass of
the system may then again easily surpass the Chandrasekhar limit.

In addition to the problems of the \sch{} scenario identified in detailed
numerical simulations, \citet{Taubenberger2013} presented compelling arguments
against any model which attempts to explain the \sndc{}-like objects by
radioactivity alone. Ignoring the detailed processes which could lead to the
production of very high nickel masses, they showed that the bolometric light
curve of an explosion with a nickel mass of $M_{\mathrm{Ni}} =
1.6\,\mathrm{M}_{\sun}$ and an ejecta mass consistent with the observed
kinematics, $M_{\mathrm{ejecta}} = 3.0\,\mathrm{M}_{\sun}$, as suggested by
\citet{Hachinger2012}, matches the observed peak luminosity of \sndc{} but
fails to reproduce the tail region owing to the increased trapping of $\gamma$
and optical radiation. Additionally, the recent study by \citet{Dhawan2016},
which aims at deriving nickel masses from infrared data, did not find any
indications for extraordinarily high nickel masses for the \sch{} objects
SN~2007if and \sndc{}.

A completely different view of the problem of \sch{} \snias{} is taken by the
interaction scenario \citep[see specifically][]{Hachinger2012,
Taubenberger2013}. Here, no extraordinarily high nickel masses are required.
Instead, an additional energy source in the form of interactions between the
ejecta and some form of CSM is invoked, which, together with the energy release
due to radioactivity, is suggested to power the light curve of \sch{} SNe, very
similar to what is discussed in the context of other super-luminous supernovae
\citep[e.g.][]{Ofek2007,Chevalier2011}. In this scenario, kinetic energy of the
supernova ejecta is converted into radiation and thermal energy in shock
heating processes during the interaction with the CSM. Already tapping a small
fraction of the kinetic energy pool of the SN ejecta could be sufficient to
explain the intense luminosities of \sch{} objects. However, the plausibility
of the interaction scenario as the mechanism underlying \sch{} objects is
challenged by the lack of clear and unambiguous interaction signatures, such as
narrow emission lines, as in the case of Type IIn supernovae, in the
observational data.

Despite this, \citet{Taubenberger2013} could construct a simple toy model with
which the tail light curve of \sndc{} could be well reproduced simply by
assuming an overall increase in ejecta density, thought to be the result of
shock compression processes.  However, being a pure radiative transfer
investigation no definitive statement about the peak of the light curve and
whether ejecta--CSM interaction could provide the requested luminosity boost
could be made. 

\section{Models and used methods}
\label{sec:model_numerics}

In this work, we consider ejecta of fairly normal \snia{} explosions and embed
them into a compact and dense carbon-oxygen CSM environment. Using a
sophisticated radiation hydrodynamical code, we follow the evolution of these
configurations and examine the exact consequences of the interaction between
the SN ejecta and the CSM. In particular, we focus on kinematic changes within
the ejecta and the effects on the optical light curve. 

In the choice of model parameters, we draw inspiration from the investigation
of \citet{Taubenberger2013}. Relying on similar parameters, we may, in addition
to studying the generic consequences of ejecta--CSM interaction, also examine
whether such a scenario provides a viable explanation for \sch{} SNe. By
construction, the particular flavour of ejecta--CSM interaction studied in this
work addresses the common criticism voiced against the interactions scenario in
this context, namely the absence of any clear unambiguous interaction
signatures in the observational data of \sch{} objects. With the restriction to
compact, carbon and oxygen rich CSM configurations, the non-detection of narrow
hydrogen or helium emission lines would be a natural consequence of the absence
of these elements in the CSM. Due to the small extent of the CSM, the
interaction phase, during which the ejecta continuously sweeps up the material,
would be quite brief. Consequently, other interaction signatures, such as high
UV fluxes, would be restricted to the early phases of the supernova evolution,
which are typically not very well observed. Such a circumstellar environment
could emerge in the context of the classical double degenerate scenario
\citep{Webbink1984}, in which the complete disruption of the secondary WD
during the merger does not induce an explosion but eventually the ensuing
accretion process of the material onto the primary WD \citep[c.f][]{Yoon2007}.
More details on the link between this `slow-merger' scenario and the CSM
configurations investigated here are provided in Section \ref{sec:csm_origin}.
In the following, we describe our input model and briefly review the numerical
tool with which the calculations are performed, before presenting our results.

\subsection{Ejecta--CSM Models}
\label{sec:input_model}

\citet{Taubenberger2013} could successfully reproduce the tail light curve of
\sndc{} in a pure radiative transfer calculation using a simple toy model,
referred to as \sndctail{}, mimicking the effects of ejecta--CSM interaction.
The model was constructed to reflect the situation of a Chandrasekhar mass
explosion, having produced $1.0\,\mathrm{M}_{\sun}$ nickel, interacting with a
pure carbon-oxygen CSM with a mass of $M_{\mathrm{CSM}} \approx
0.65\,\mathrm{M}_{\sun}$ . Since hydrodynamic and radiation hydrodynamic
effects have been neglected, the consequences of ejecta--CSM interaction are
crudely taken into account by a global increase of the density (which is set up
as an exponential profile). 

\begin{figure}
  \centering
  \includegraphics[]{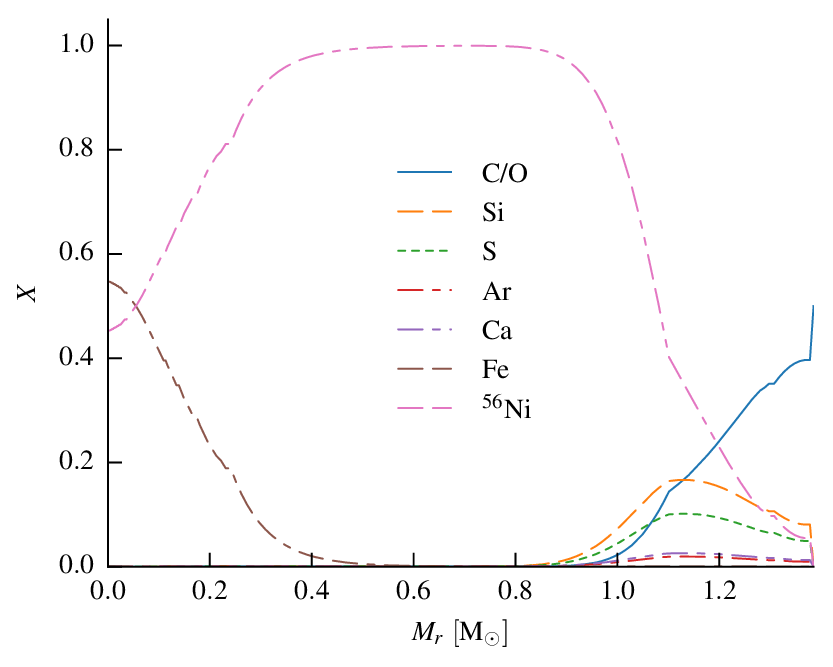}
  \caption{Illustration of the abundance structure of the ejecta model
    \texttt{m100101}. The mass fraction $X$ of a number of important elements
  and isotopes is shown.}
  \label{fig:ejectaabundances}
\end{figure}

Our numerical calculations are inspired by the \sndctail model, but we aim at a
self-consistent treatment, in particular of the interaction phase and its
consequences.  For this purpose, we rely on the suite of explosion models
produced by \citet{Woosley2007} to initiate the ejecta in our simulations. In
these explosion calculations, the ejecta structure and evolution has been solved
self-consistently in accordance with the energy released in the thermonuclear
burning leading to a pre-defined ejecta composition. In particular, we rely on
the model \texttt{m100101} of the series which closely reflects the parameters
adopted by \citet{Taubenberger2013}. The main characteristics of this ejecta
model are summarised in Table \ref{tab:ejectamodel} and its detailed elemental
composition is illustrated in Figure \ref{fig:ejectaabundances}.
\begin{table}
  \centering
  \caption{Main properties of the ejecta model \texttt{m100101}, produced by
  \citet{Woosley2007} and used in our study.}
  \label{tab:ejectamodel}
  \begin{tabular}{lr}
    \hline
    Quantity & Value \\
    \hline
    $M_{\mathrm{tot}}$ & $1.39\,\mathrm{M}_{\sun}$\\
    $M_{\mathrm{Ni}}$  & $1.00\,\mathrm{M}_{\sun}$\\
    $M_{\mathrm{Fe}}$  & $0.10\,\mathrm{M}_{\sun}$\\
    $M_{\mathrm{C/O}}$ & $0.18\,\mathrm{M}_{\sun}$\\
    $M_{\mathrm{IME}}$ & $0.10\,\mathrm{M}_{\sun}$\\
    $E_{\mathrm{kin}}$ & $1.44\times 10^{51}\,\mathrm{erg}$\\
    \hline
  \end{tabular}
\end{table}

\begin{figure}
  \centering
  \includegraphics{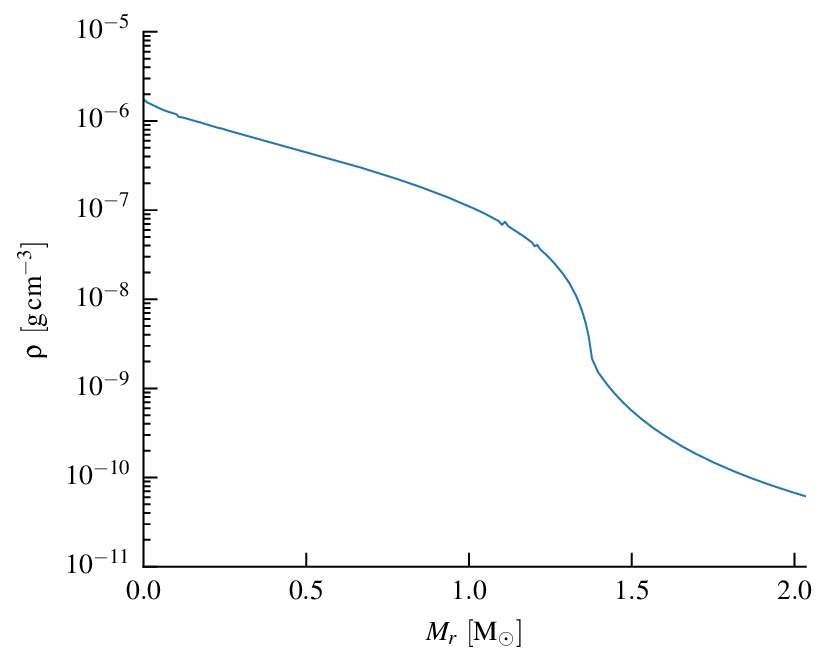}
  \caption{Density profile of the ejecta--CSM model which is used as the starting point
    for our \stella simulations, $t_{\mathrm{exp}} = \SI{e4}{s}$ after explosion.}
  \label{fig:ejectacsmmodel}
\end{figure}

We embed the ejecta into a dense CSM of $M_{\mathrm{CSM}} = 0.64\,\mathrm{M}_{\sun}$,
again very similar to the choices of \citet{Taubenberger2013}. This material is
assumed to be free of hydrogen and helium but rich in carbon and oxygen and of
uniform composition. The CSM composition can be described by
\begin{align}
  X_i = \begin{cases}
    f_{Z} X_{i,\sun} & \quad i = 7, i > 8\\
    \left(1 - f_{Z} X_{7,\sun} - f_{Z} \sum_{i>8} X_{i,\sun}\right) / 2 & \quad i = 6,8\\
    0 & \quad i < 6
  \end{cases}
  \label{eq:composition_metalicity_boosted}
\end{align}
with $X_i$ denoting the mass fraction of the element with proton number $i$.
The scaling factor $f_Z$ is a measure for the amount of non carbon-oxygen
material in the CSM relative to solar abundances. We start our investigation
with $f_Z=1$, i.e.\ using solar abundances for all elements up to $Z=28$,
except for hydrogen, helium, lithium, beryllium, boron (which are all zero) and
carbon and oxygen (which make up the remaining mass in equal parts). We further
assume that the CSM is initially at rest and follows a power-law density
profile, i.e.\
\begin{equation}
  \rho_{\mathrm{CSM}} \propto r^{-p}.
  \label{eq:wind_density}
\end{equation}
with $p = 2$, extending up to an outer radius $R_{\mathrm{CSM}} =
\SI{1.3e14}{cm}$, which coincides with the outer boundary of the computational
domain. The resulting density structure of the complete model is shown in
Figure \ref{fig:ejectacsmmodel}.

\subsection{Numerical Method}
\label{sec:stella}

To accurately track the shock-heating and compression processes together with
the simultaneous transport of radiative energy generated in the ejecta--CSM
interaction and produced by radioactivity, a radiation hydrodynamical treatment
is required. For this purpose, we rely on the numerical code \stella
\citep{Blinnikov1993,Blinnikov1998,Blinnikov2006}. A number of design features
render this code an ideal choice for studying interacting supernovae, in particular
its fully implicit treatment of the radiation hydrodynamical problem and the
Lagrangian description of fluid dynamics. These characteristics are particularly
important when dealing with the strong forward and reverse shocks which are
expected to form at the ejecta--CSM interface. A Lagrangian treatment, in which
the computational discretisation naturally adapts to the local fluid flow, is
ideal for accurately tracking and resolving the large density contrasts in the
vicinity of the shocks. 

To address radiation hydrodynamical problems, it is often advantageous to rely
on fully implicit treatments due to discrepancies between the characteristic
flow and radiative time scales \citep[see for example][for a discussion of the
radiation hydrodynamical coupling]{Lowrie1999,Sekora2010}. This concern is
particularly relevant for the type of problem investigated in this work. Due to
the high kinetic energy of the ejecta material, intense heating is expected to
occur at the shocks. The characteristic time scales of the resulting cooling
processes may easily become much smaller than the hydrodynamic scales. This
typically poses problems when relying on explicit time-stepping
procedures\footnote{In the context of finite volume schemes, this is often
referred to as the stiff source term regime \citep[e.g.][]{Leveque2002}.}
since all relevant time scales have to be resolved.

Given \stella's advantages, it is ideally suited to study interacting supernovae
and has been used for this purpose on numerous occasions
\cite[e.g.][]{Blinnikov2010,Baklanov2015,Sorokina2016a}.  Since \textsc{Stella}
is a well-established technique, we only highlight some key aspects of the
implemented physical processes and numerical techniques here.  More details may
be found in \citet{Blinnikov1993,Blinnikov1998,Blinnikov2006}.  \textsc{Stella}
operates on a one-dimensional, spherically symmetric, Lagrangian mesh and solves
the full radiation hydrodynamical problem implicitly, including
multi-group radiative transfer in the co-moving frame. The equations are
discretised using finite differences for spatial and frequency derivatives,
resulting in a large system of ordinary differential equations in time. This
flavour of the method of lines \citep[c.f.][]{Oran1987} is solved by using the
multi-step predictor-corrector schemes of \citet{Gear1971} and
\citet{Brayton1972}. 

When calculating the radiation--matter coupling, \textsc{Stella} takes
contributions due to electron scattering, inverse bremsstrahlung,
photoionization and interactions with atomic lines into account. For the last
process, the expansion opacity treatment of \citet{Friend1983,Eastman1993} is
adopted \citep[see details in][]{Blinnikov1998}. In addition to the multi-group
radiative transfer treatment, a single-group $\gamma$-transport scheme is
included to account for the energy deposition due to the main radioactive decay
chain in \snias{}, \ce{^{56}Ni -> ^{56}Co -> ^{56}Fe} \citep{Blinnikov2006}.

The radiative transfer treatment just detailed is less sophisticated than
in dedicated pure radiative transfer and spectral synthesis approaches, such as
\textsc{Artis} \citep{Kromer2009}, \textsc{Cmfgen}
\citep{Hillier1998,Hillier2012}, \textsc{Phoenix} \citep{Hauschildt2004},
\textsc{Sedona} \citep{Kasen2006} or \textsc{SuperNu}
\citep{Wollaeger2013,Wollaeger2014}. Moreover, the \stella calculations are
currently restricted to one-dimensional spherically symmetric setups. Both
compromises are owed to \stella's fully implicit design and its ambition to
solve the fully coupled radiation hydrodynamical equations, properties which
are crucial for the type of problem investigated here (but see also discussion
in Sections \ref{sec:discussion_geometry} and \ref{sec:discussion_linelist}).

\subsection{Numerical Setup}
\label{sec:num_setup}

All results described in the following have been obtained in simulations with
the same numerical parameters (exceptions are explicitly stated). The input
model has been discretised on a Lagrangian mesh with 358 shells, with the
ejecta model occupying the first half and with the remaining cells describing
the CSM configuration. In addition to the spatial discretisation, the frequency
space between \parnumin{} and \parnumax{} has been subdivided into 300
logarithmically spaced bins. This range corresponds to the wavelength regime
\parlammin{} and \parlammax{}. All simulations are started $t_{\mathrm{exp}} =
\SI{e4}{s}$ after explosion (c.f.\ Section \ref{sec:input_model}). An overview
of all the calculations performed and presented in this work is provided in
Table \ref{tab:models_overview}.
\begin{table*}
  \centering
  \caption{Overview of the different ejecta-CSM configurations investigated and
    presented in this work. The number of radial shells $N_{\mathrm{shells}}$,
    the total mass of the CSM $M_{\mathrm{CSM}}$ and its composition in terms
    of the parameter $f_Z$ (see Equation
    \ref{eq:composition_metalicity_boosted}) are provided. Also, the starting
    time of the calculation, $t_{\mathrm{exp}}$, and the outer radius,
    $R_{\mathrm{out}}$ of the computational domain are indicated. The latter
    corresponds to the initial outer edge of the CSM envelope except for the
    pure ejecta model.  Here, $R_{\mathrm{out}}$ denotes the radius of the
  outer edge of the ejecta.}
  \begin{tabular}{lcccccr}
    \hline
    model name & $N_{\mathrm{shells}}$ & $M_{\mathrm{CSM}}$ [$\mathrm{M}_{\sun}$] & $f_{Z}$ & $t_{\mathrm{exp}}$ [s] & $R_{\mathrm{out}}$ [cm] & notes\\
    \hline
    \modnocsm{}        & 179 & 0    & N/A & $10^4$ & $\SI{2.53e13}{}$ & bare ejecta model \\ 
    \modfid{}          & 358 & 0.64 & 1   & $10^4$ & $\SI{1.28e14}{}$ & fiducial interaction model \\ 
    \modztwo{}         & 358 & 0.64 & 2   & $10^4$ & $\SI{1.28e14}{}$ & \\ 
    \modzfive{}        & 358 & 0.64 & 5   & $10^4$ & $\SI{1.28e14}{}$ & \\ 
    \modztwenty{}      & 358 & 0.64 & 20  & $10^4$ & $\SI{1.28e14}{}$ & best-fit model concerning light curve shape \\ 
    \modzhundred{}     & 358 & 0.64 & 100 & $10^4$ & $\SI{1.28e14}{}$ & \\ 
    \modmassivecsm{}   & 358 & 1.28 & 1   & $10^4$ & $\SI{1.28e14}{}$ & \\ 
    \modfidearly{}     & 358 & 0.64 & 1   & $10^3$ & $\SI{1.29e14}{}$ & \\ 
    \modfidearlier{}   & 358 & 0.64 & 1   & $10^2$ & $\SI{1.31e14}{}$ & \\ 
    \modfidesmall{}    & 358 & 0.64 & 1   & $10^2$ & $\SI{5.21e13}{}$ & \\ 
    \modfidesmaller{}  & 358 & 0.64 & 1   & $10^2$ & $\SI{2.59e13}{}$ & \\ 
    \modfidesmallest{} & 358 & 0.64 & 1   & $10^2$ & $\SI{1.29e13}{}$ & \\ 
    \hline
  \end{tabular}
  \label{tab:models_overview}
\end{table*}

\section{Calculations and Results}
\label{sec:results}

By following the radiation hydrodynamical evolution of the input model described
in Section \ref{sec:input_model} with \stella, we aim at exploring the
consequences of thermonuclear explosions occurring within a dense and compact
CSM (see sections \ref{sec:evolution_general} and \ref{sec:lc_general}).  In
addition, we examine whether the ensuing ejecta--CSM interaction may explain
the peculiar properties of \sch{} objects by performing a detailed comparison
with \sndc{} in Section \ref{sec:cmp_09dc}.

\subsection{General Model Evolution}
\label{sec:evolution_general}

\begin{figure}
  \centering
  \includegraphics{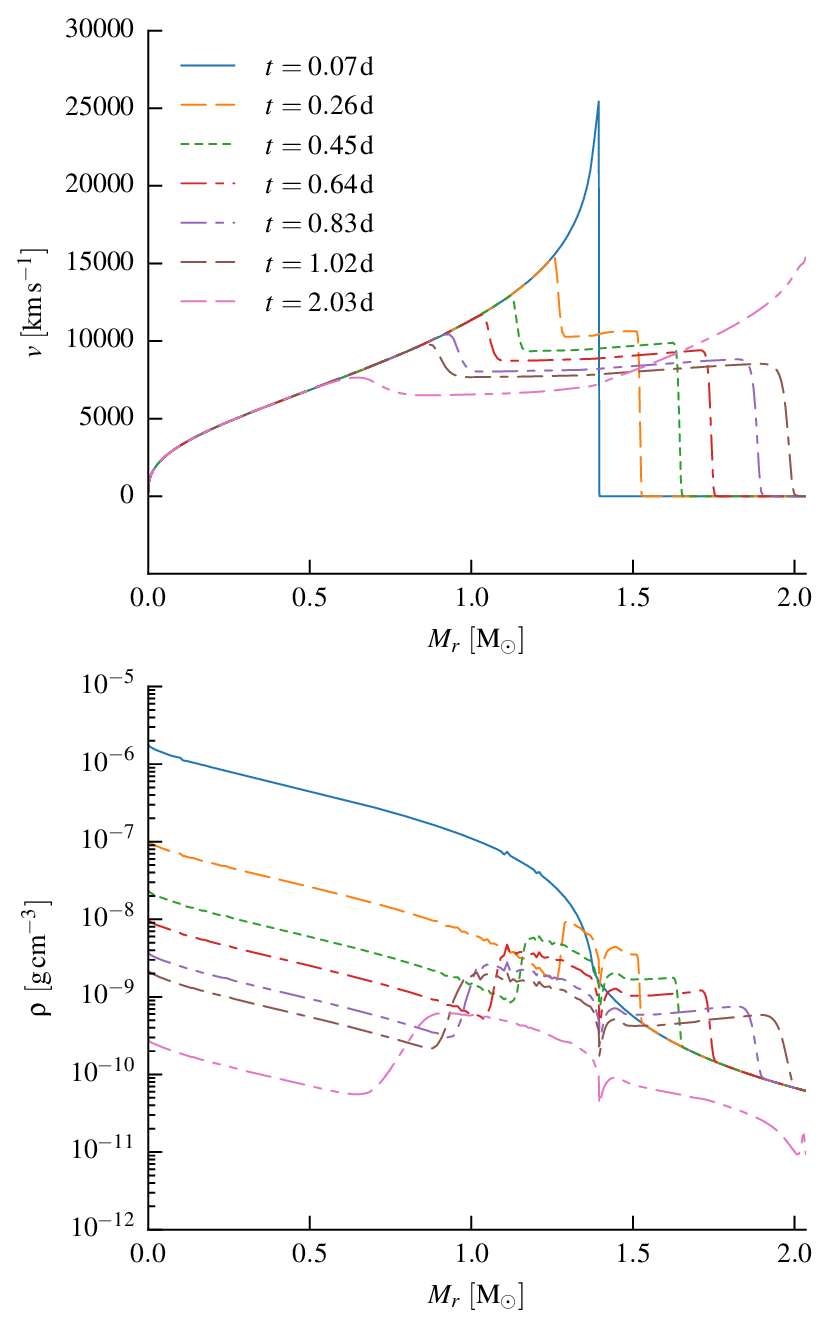}
  \caption{Velocity (upper panel) and density (lower panel) evolution of the
    interacting \snia{} model \modfid{}.  Different snapshots are shown. The
    reverse shock steadily propagates inwards in mass-space, compressing and
    decelerating the ejecta material. The same qualitative behaviour is
  observed in all other interaction calculations performed in this work as
well.}
  \label{fig:ejectacsmevolution}
\end{figure}

\begin{figure}
  \centering
  \includegraphics{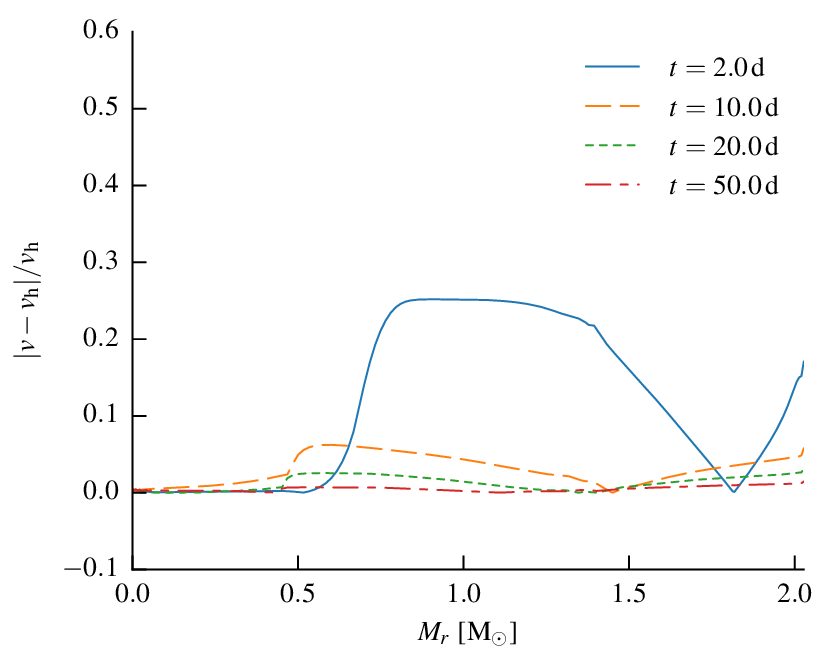}
  \caption{Relative deviation from homology in the ejecta--CSM interaction model \modfid{}
  for a number of snapshots after the forward shock has swept up the entire
  CSM.  Already at time $t = \SI{10}{d}$, the largest deviations are less then
  ten per cent. Note that the time since explosion, $t_{\mathrm{exp}}$, is used
  for the determination of the homologous expansion velocity $v_{\mathrm{h}} = r /
  t_{\mathrm{exp}}$.}
  \label{fig:modelhomology}
\end{figure}

At the ejecta--CSM interface, fast moving ejecta material collides with the
initially inert CSM. As a consequence, strong forward and reverse shocks emerge
from the interface at the start of the calculations. The forward shock ploughs
through the CSM and continuously sweeps up material while the reverse shock
propagates into the ejecta in a Lagrangian sense, i.e.\ penetrating ever smaller
mass shells\footnote{Note that the radius of the reverse shock still
continuously increases with time.}. By this process the reverse shock continuously
compresses, heats and decelerates the ejecta material as illustrated in Figure
\ref{fig:ejectacsmevolution}.
Due to the compact configuration of the CSM, the forward shock quickly sweeps
over the entire ambient material, after which the continuous conversion of
kinetic into thermal energy subsides. This occurs at about $t \approx \SI{1.2}{d}$, after which
the model quickly readjusts and establishes a homologously expanding flow as
illustrated in Figure \ref{fig:modelhomology}.
We note that, unless explicitly stated otherwise, we do not express time during
the model evolution relative to the explosion point, but apply a crude `light
travel correction', accounting for the continuous expansion of the Lagrangian
numerical grid. In particular, for a snapshot at $t_{\mathrm{exp}}$ after
explosion, at which the outer edge of the computational domain is at
$r_{\mathrm{out}}$, this `observer time' is given by
\begin{equation}
  t = t_{\mathrm{exp}} - \frac{r_{\mathrm{out}}}{c}.
  \label{eq:observer_time}
\end{equation}
Here, the speed of light $c$ appears.

\subsection{Light Curve}
\label{sec:lc_general}

\begin{figure}
  \centering
  \includegraphics{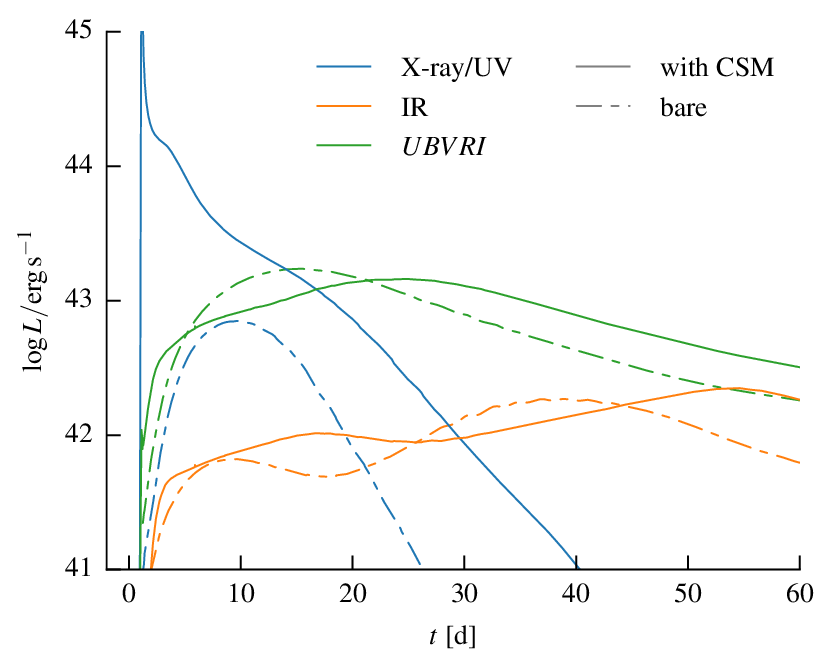}
  \caption{Comparison between the different broad-band light curves of the
    interacting \snia{} model (\modfid{}, solid lines) and a corresponding calculation in which
    only the bare ejecta has been considered (\modnocsm{}, dashed lines). Note that the light
curves have been determined by applying simple cuts to the SED (c.f.\ Table
\ref{tab:broadbandcuts}).}
  \label{fig:modellightcurve}
\end{figure}

During the ejecta--CSM interaction phase, very high temperatures are induced in
the shock region, leading to intense emission of radiation from the shock front.
Due to the high temperatures, the radiation is mostly emitted in the extreme
ultraviolet and X-ray bands. This leads to an intense short-duration flash in
this regime as seen in Figure \ref{fig:modellightcurve}.
The light curves shown there are determined by applying cuts to the SED of the
emergent radiation field as detailed in Table
\ref{tab:broadbandcuts}.
\begin{table}
  \centering
  \caption{Cuts in wavelength space which have been applied to determine the
    model light curves in the different regimes from the emergent SED in the
    \stella calculations. In addition, the number of bins of the
      logarithmically spaced frequency grid (see Section \ref{sec:num_setup}),
      which discretize the different regimes, $N_{\nu}$, are indicated.}
  \label{tab:broadbandcuts}
  \begin{tabular}{lrrr}
    \hline
    regime & $\lambda_{\mathrm{min}}$ [\AA] & $\lambda_{\mathrm{max}}$ [\AA] & $N_{\nu}$ \\
    \hline
    X-ray/UV & 1 & 3250 & 224\\
    $UBVRI$ & 3250 & 8900 & 28\\
    IR & 8900 & 49000 & 48 \\
    \hline
  \end{tabular}
\end{table}
Only a small fraction of the radiation in the high energy regime is reprocessed
and shifted into the optical bands by interactions in the un-shocked material.
This behaviour leads to a first bump in the $UBVRI$ light curve as seen in Figure
\ref{fig:modellightcurve}.

After the forward shock has reached the outer CSM edge, the shock heating
ceases, the shocked material gradually cools and the nickel decay becomes the
dominant process in shaping the light curve. Compared to a `bare' ejecta
calculation (\modnocsm{}), however, an increased trapping of $\gamma$ and
optical radiation is observed. This is caused on the one hand by the additional
material which has been piled upon the ejecta and on the other hand by the
density enhancement due to the shock compression. This leads to a longer rise
time, a broadening of the peak and to a slower decline in the tail regime as
highlighted by the comparison with the results of the `bare' ejecta
calculation \modnocsm{}, which are included in Figure \ref{fig:modellightcurve}. However,
due the small extent of the CSM configuration in our calculations, the ensuing
ejecta--CSM interaction does not lead to a significant boost in the optical
light curve around maximum.

\subsection{Comparison with \sndc}
\label{sec:cmp_09dc}

As outlined above in the generic description of the model evolution, the
ejecta--CSM interaction induces a number of features which qualitatively match
some of the defining properties of the class of \sch{} objects.  For a more
detailed assessment of the validity of the proposed scenario, we perform in the
following sections a detailed quantitative comparison between the model
properties and the observed data of the prototype of the class, \sndc{}.

\subsubsection{Line Velocities}

\begin{figure}
  \centering
  \includegraphics{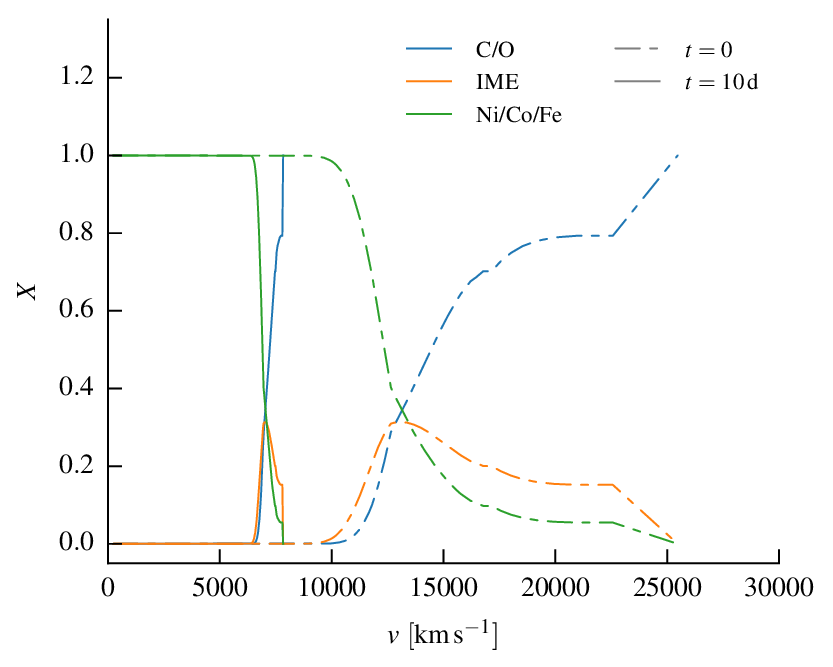}
  \caption{Comparison of the initial distribution of a number of selected
    elements in the ejecta material to that after the main ejecta--CSM
    interaction phase in \modfid{}.  In particular, the situation at $t =
    \SI{10}{d}$ is shown. By that time, the reverse shock has efficiently
  decelerated most of the ejecta material.}
  \label{fig:elementalevolution}
\end{figure}

Among the  striking properties of \sndc{} are its peculiar line velocities,
which are consistently lower than in normal \snias{} \citep[e.g.][]{Taubenberger2011}.
In the investigated model, the strong reverse shock quickly and efficiently
decelerates the ejecta material. This leads to an elemental distribution in
velocity space which is significantly compressed compared to that of a typical
\snia{}. Figure \ref{fig:elementalevolution} illustrates this behaviour in \modfid{} by
comparing the initial stratification of the ejecta material (representative of the situation in a
typical \snia{}) with the situation after the interaction phase, roughly at the
time when homology is nearly re-established.
Already at this early time, i.e.\ $t = \SI{10}{d}$, all intermediate-mass
elements (IME) are confined to a very small low-velocity region, ranging from about
$\SI{6000}{km.s^{-1}}$ to $\SI{8000}{km.s^{-1}}$. This distribution is
compatible with the line velocities determined for \sndc: in particular
\citet{Taubenberger2013} established that the Si~II~$\lambda 6355$ line
velocity lies between $\SI{6000}{km.s^{-1}}$ and $\SI{9000}{km.s^{-1}}$.

One should emphasise that this consideration has solely been based on the
elemental distribution in velocity space, without taking the details of the line
formation process into account. Ideally, the line velocities would be validated
by comparing synthetic with observed spectra. However, the SEDs provided by
\stella are much too coarse for this purpose. Nevertheless, a necessary
condition for the appearance of certain line velocities in the observed spectral
features is the location of the corresponding elements in the appropriate
velocity regime.

\subsubsection{$UBVRI$ Light Curve}

\begin{figure}
  \centering
  \includegraphics{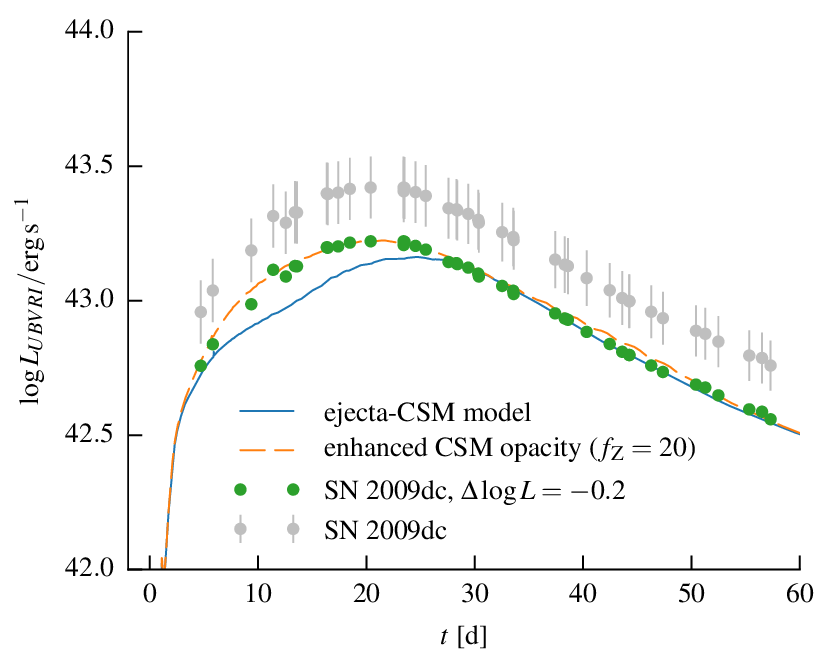}
  \caption{Comparison between the synthetic $UBVRI$ light curve of the
    interacting model (\modfid{}, solid blue line) and the corresponding
    observed luminosities for \sndc{} (grey symbols). The displayed uncertainty
    ranges are due to distance and reddening. To highlight the good match
    between the synthetic light curve shape and the observed evolution, a
    comparison with the observed data shifted by $\Delta \log
    L/\mathrm{erg\,s^{-1}} = -0.2$ (green symbols) is included as well. An
    even better match to the light curve shape is achieved when enhancing the
    reprocessing efficiency in the CSM (by artificially boosting the
    content of non-carbon-oxygen material in the CSM, see Section
    \ref{sec:peak}). The synthetic $UBVRI$ light curve of one such calculation,
    with $f_Z = 20$, (see Equation \ref{eq:composition_metalicity_boosted}) is
    included (\modztwenty{}, dashed orange line)} 
  \label{fig:lc_cmp_09dc_shift}
\end{figure}

As highlighted in Section \ref{sec:lc_general} and illustrated in Figure
\ref{fig:modellightcurve}, the interaction between ejecta and CSM leads to a
slower rise of the $UBVRI$ light curve, a broadening of its overall
shape and a slower decline -- all properties which have been observed in
\sndc{}. However, when directly comparing the synthetic $UBVRI$ light curve with
the observed data of \sndc{}, as done in Figure \ref{fig:lc_cmp_09dc_shift}, the
missing overall luminosity boost becomes apparent.
In particular, the \stella model light curve is during all phases too dim by
about $0.2\,\mathrm{dex}$. This discrepancy goes beyond the uncertainties set
by the distance determination and adopted reddening prescription for \sndc{}
\citep[see][]{Taubenberger2011}. Despite this mismatch on absolute scales, the
interaction model reproduces the observed light curve shape remarkably well.
This is illustrated in the comparison shown in Figure
\ref{fig:lc_cmp_09dc_shift}, in which a global, artificial shift of
$0.2\,\mathrm{dex}$ has been applied to the \sndc{} observations. This is
comparable to the assumption made by \citet{Kamiya2012} that there be no host
galaxy extinction\footnote{Note that the complete absence of any host
extinction is ruled out by the detection of interstellar sodium lines at the
redshift of the host of \sndc{} \citep[e.g.][]{Taubenberger2011}.}.

\subsubsection{Addressing the Peak}
\label{sec:peak}

The largest discrepancy in the light curve shape between model and observations
manifests around the peak (see Figure \ref{fig:lc_cmp_09dc_shift}). During this
phase, the model exhibits two pronounced plateaus. The initial steep rise
represents the optical afterglow of the interaction phase which is superimposed
on the normal typical light curve evolution due to radioactive heating. As a
consequence of the shock compression, the radioactive signal is delayed. In the
peak region, the synthetic light curve is also lacking more luminosity than in
the other regimes.

The ejecta--CSM interaction nevertheless generates a substantial amount of
radiative energy. However, due to the high shock temperatures this energy is
mostly released in the X-ray and far UV regime. Only a small fraction is
reprocessed by interacting in the CSM, thus producing the optical afterglow seen
as the first plateau in the light curve.

\begin{figure}
  \centering
  \includegraphics{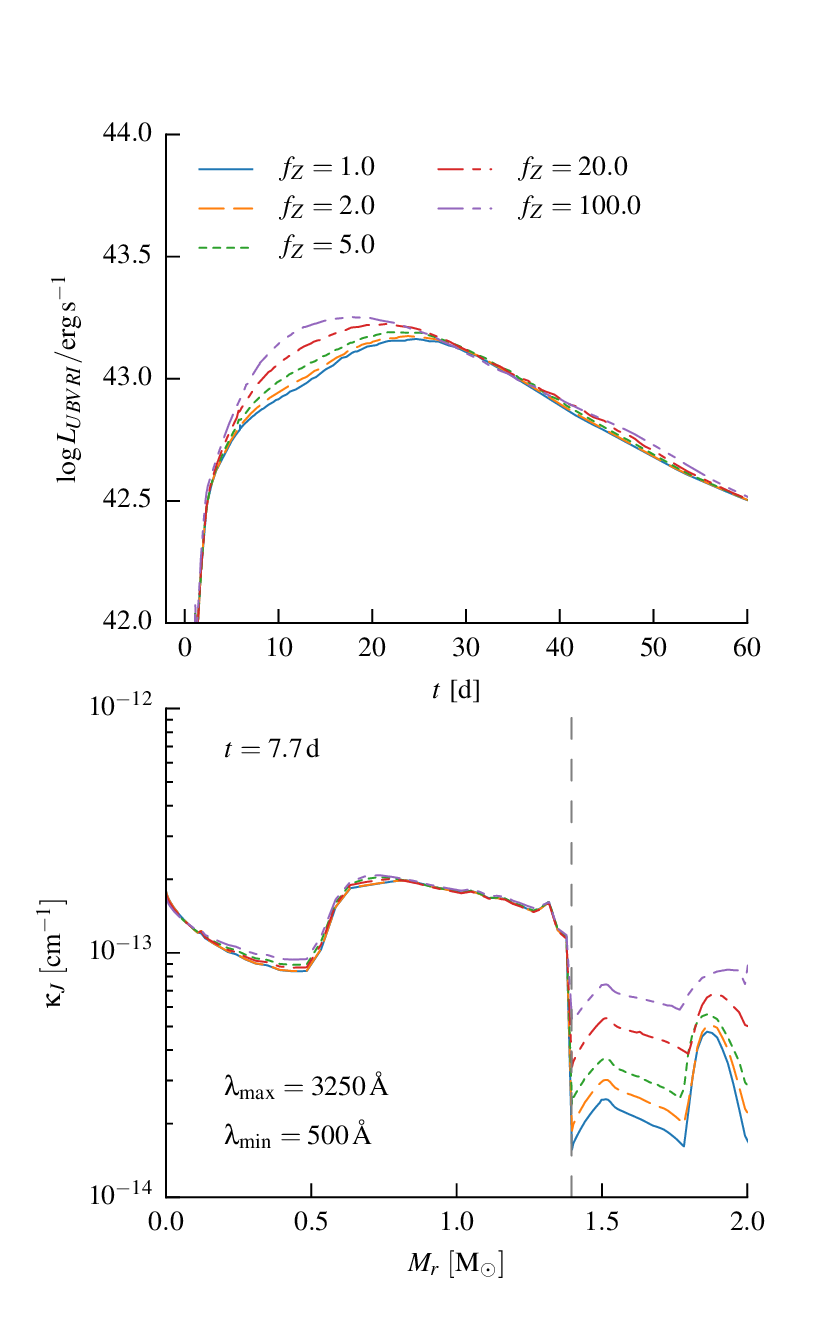}
  \caption{Consequences of increasing the opacity in the CSM by changing its
    composition according to Equation
    (\ref{eq:composition_metalicity_boosted}). In particular, the models
    \modfid, \modztwo, \modzfive, \modztwenty and \modzhundred are shown.  The
    composition change leads to a stronger redistribution of high energy
    radiation generated during the ejecta--CSM interaction into the optical
    $UBVRI$ bands around peak (upper panel). In the bottom panel, the increase
    in opacity when changing $f_{Z}$ is illustrated in terms of the
    absorption-mean $\kappa_{J}$ (see Equation
    \ref{eq:absorption_mean_opacity}), evaluated in the range
    $\SI{500}{\angstrom} \le c/\nu \le \SI{3250}{\angstrom}$.}
  \label{fig:lc_opacity_boosted_metalicity}
\end{figure}

This leads to the conclusion that a luminosity boost may still be achieved
within the proposed interaction scenario, provided that the re-processing
efficiency can be enhanced. To demonstrate this, we artificially boost the
opacity in the CSM. A simple way to achieve this within the \stella framework
to increase the amount of non-carbon-oxygen material. For
simplicity, we just scale the solar abundances for all elements heavier than
$Z=8$ (including nitrogen) and set the remainder to carbon and oxygen in equal
parts. Thus, we use a scaling parameter $f_Z$ different from one in Equation
(\ref{eq:composition_metalicity_boosted}). Figure
\ref{fig:lc_opacity_boosted_metalicity} illustrates the consequences of
boosting the non-carbon-oxygen material according to the prescription
(\ref{eq:composition_metalicity_boosted}).
The increase in opacity when using higher metal contents in the CSM is shown in
terms of the corresponding absorption-mean \citep[c.f.][\S 82]{Mihalas1984}
\begin{equation}
  \kappa_J = \frac{\int \mathrm{d}\nu \kappa_{\nu} J_{\nu}}{\int
    \mathrm{d}\nu J_{\nu}},
  \label{eq:absorption_mean_opacity}
\end{equation}
involving a frequency integration of the absorption opacity $\kappa_{\nu}$,
weighted by the mean intensity $J_{\nu}$.  As the opacity and thus the
reprocessing efficiency in the CSM increases, the $UBVRI$ luminosity in the
peak region is steadily boosted. Figure \ref{fig:lc_cmp_09dc_shift} also
compares the synthetic light curve of the $f_Z = 20$ calculation,
\modztwenty{}, with the \sndc{} observations (still shifted by
$0.2\,\mathrm{dex}$), now showing a perfect match in the light curve shape.

We emphasize that this exploration only serves to illustrate the general
consequences of having a high reprocessing efficiency for energetic radiation
in the CSM. This significantly improves the observable optical light curve of
the interacting model. We stress, however, that these results should not
interpreted as indications for the presence of a particular type of CSM
with a high amount of non-carbon-oxygen material. In fact,
\citet{Childress2011} and \citet{Khan2011} suggest low-metallicity
environments for \sch{} events. Instead, we consider the CSM
composition change a simple realisation of enhancing the opacity in the
\stella framework and as a potential representation of interaction processes
which are not fully incorporated in the simplified radiative transfer treatment
adopted in \stella. One such shortcoming is the small number of atomic line
transitions taken into account. A more detailed discussion of this point will
follow in Section \ref{sec:discussion_linelist}.

\subsubsection{Addressing the Tail}
\label{sec:09dc_cmp_tail}

The experiments with an enhanced opacity in the CSM described in the previous
section nicely lead to a luminosity boost around the peak. However, the
post-maximum decline and the tail of the light curve are largely unaffected by
these changes in the CSM composition (see top panel in Figure
\ref{fig:lc_opacity_boosted_metalicity}). But
 also during these phases, the
synthetic light curves of the interaction models are too dim and would require a
boost of $0.2\,\mathrm{dex}$ to be compatible with the
observations.

\begin{figure}
  \centering
  \includegraphics{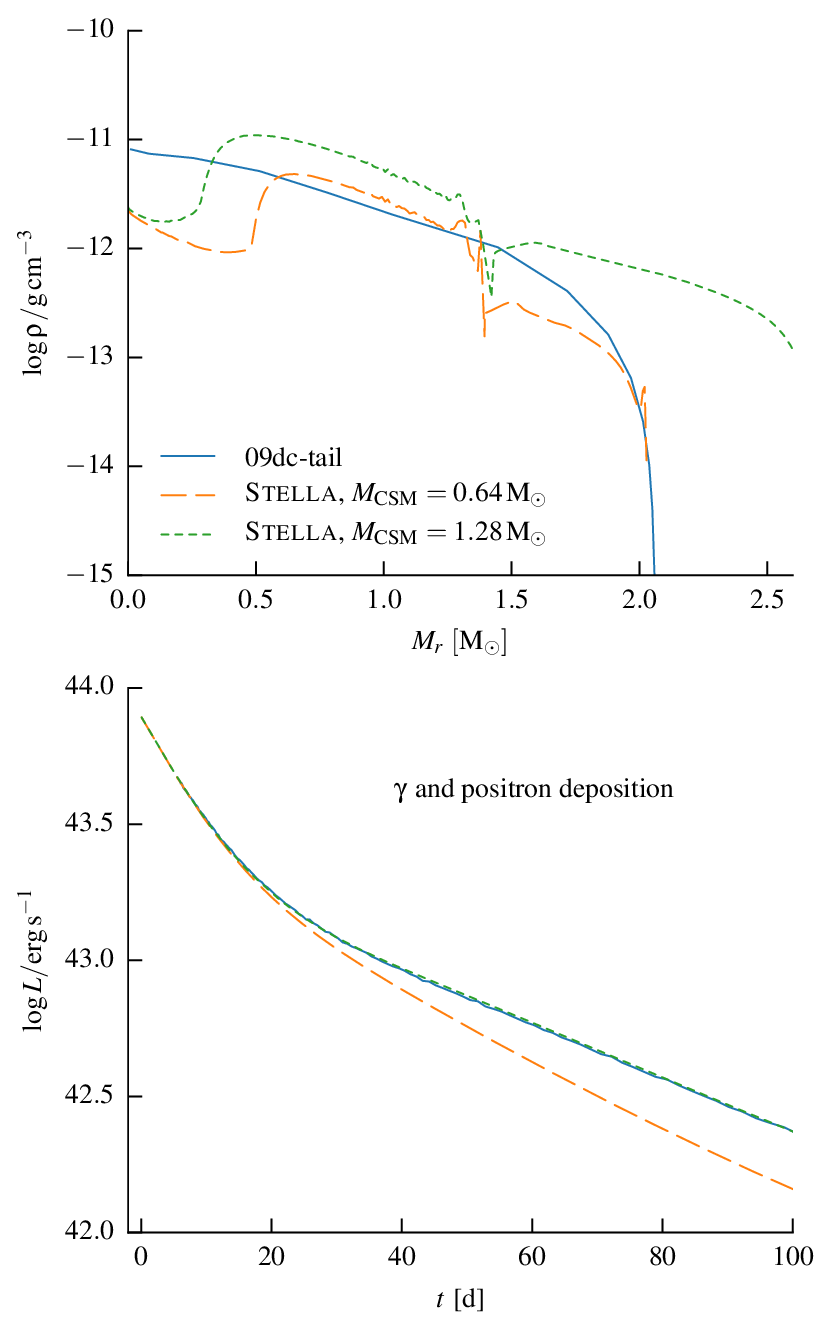}
  \caption{Comparison between the density profiles of the \stella interaction
    model (with $M_{\mathrm{CSM}} = 0.64\,\mathrm{M}_{\sun}$, \modfid) and the
    toy model \texttt{09dc-tail} by \citet{Taubenberger2013} at $t =
    \SI{10}{d}$ (upper panel). The reverse shock has stopped its propagation
    after reaching the mass shell $M_r = 0.5 \, \mathrm{M}_{\sun}$.  During the
    following evolution of the model, the density discontinuity is simply
    advected by the expanding flow. As a consequence of the differences in
    central densities, $\gamma$-ray trapping is less efficient in \stella than
    in the \artis \citep{Kromer2009} radiative transfer calculations used by
    \citet{Taubenberger2013}. The $\gamma$-deposition (lower panel) between the
    model differs by a factor of 1.5, exactly the same value by which the
    $UBVRI$ light curves are different.  For comparison, results of a \stella
    calculation with higher CSM mass, $M_{\mathrm{CSM}} =
    1.28\,\mathrm{M}_{\sun}$ (\modmassivecsm), and thus a stronger reverse shock
  are also included.}
  \label{fig:stella_09dc_tail_density_gamma}
\end{figure}

At first glance, this discrepancy in the light curve tail is unexpected given that
the ejecta--CSM interaction models have been constructed to reflect the
global properties of the toy model \sndctail{} presented by
\citet{Taubenberger2013}, as detailed in Section \ref{sec:input_model}. The
\sndctail{} toy model has been specifically designed to reproduce
the tail light curve of SN 2009dc both in shape and in absolute scale. However,
simplistic assumptions about the effects of ejecta--CSM interaction have been
adopted in the toy model (due to the lack of a self-consistent radiation
hydrodynamical treatments). More specifically, the density profile has been
globally up-scaled to approximately reflect the compression by the reverse
shock. This idealistic situation is not observed in our models, in which the
shock compression is treated self-consistently. As illustrated in Figure
\ref{fig:stella_09dc_tail_density_gamma}, the compression of the density profile
only reaches down to the mass shell $M_{r} \approx 0.5\, \mathrm{M}_{\sun}$.
The ongoing compression of the material lying above this shell has exhausted the
energy of the reverse shock. The density discontinuity is simply advected by the
expanding flow and the inner core region remains un-compressed. As a
consequence, the density within a significant fraction of the nickel core is
much lower than under the idealised assumptions of the toy-model. This leads to
a significant difference in the $\gamma$-trapping efficiency, which mainly
determines the light curve evolution past maximum. Due to the lower column
density in the \stella model, $\gamma$-radiation can escape more easily and the
$\gamma$-deposition within the ejecta is lower by roughly the same amount as in
the $UBVRI$ light curve (see lower panel of Figure
\ref{fig:stella_09dc_tail_density_gamma}).

\begin{figure}
  \centering
  \includegraphics{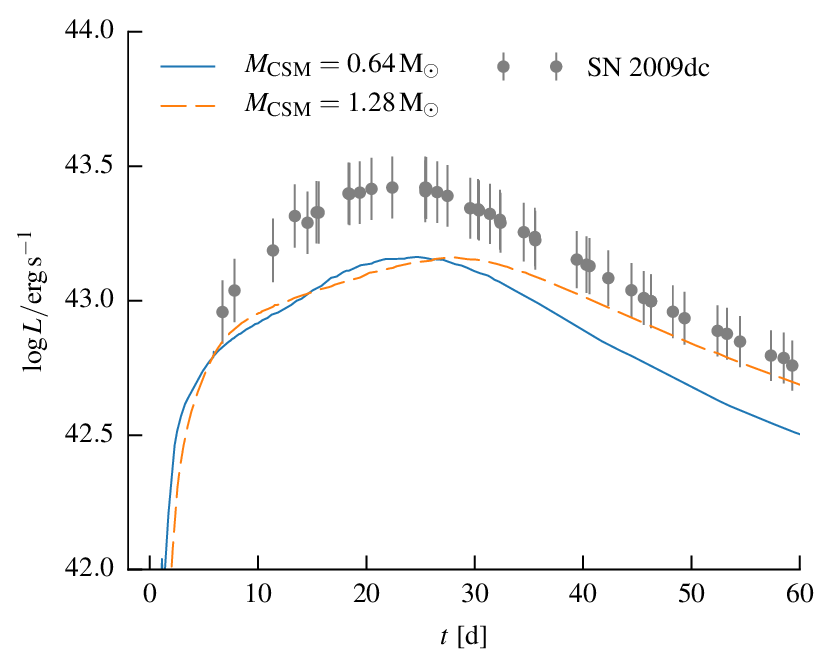}
  \caption{Comparison between the synthetic $UBVRI$ light curve of the high-CSM-mass
    \stella model \modmassivecsm and the (original) \sndc{} observations. Due to the higher
    compression of the ejecta, the light curve tail now better matches the
    observations. However, the elemental distribution in velocity space is now
    incompatible with the line velocity observations of \sndc{} (see Figure
    \ref{fig:stella_09dc_tail_highcsm_elems}). As a comparison, the light curve
    of the standard \stella interaction model, \modfid, is included as well.}
  \label{fig:stella_09dc_tail_highcsm_lc}
\end{figure}

\begin{figure}
  \centering
  \includegraphics{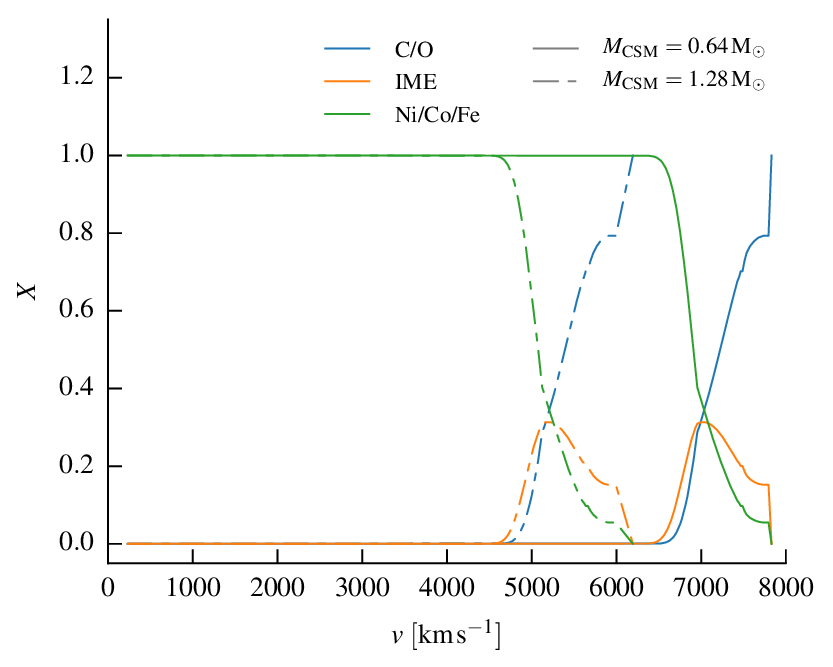}
  \caption{Final (at $t = \SI{10}{d}$ after explosion) elemental distribution
  in the normal and high-CSM-mass \stella interaction model, \modfid and
  \modmassivecsm respectively. The more powerful reverse shock in the latter
  one leads to a much stronger deceleration of the ejecta material.}
  \label{fig:stella_09dc_tail_highcsm_elems}
\end{figure}

These findings seem to indicate that a stronger compression of the nickel zone
is required to boost the tail light curve. This suspicion is confirmed by a
\stella test calculation, \modmassivecsm, in which a higher CSM mass is used ($M_{\mathrm{CSM}} =
1.28\,\mathrm{M}_{\sun}$). A stronger reverse shock is generated in this calculation,
which penetrates deeper layers, down to $M_r \approx 0.2\,\mathrm{M}_{\sun}$ and
thus leads to an overall stronger compression (see Figure
\ref{fig:stella_09dc_tail_density_gamma}).
As expected, $\gamma$-trapping is enhanced, leading to a slower decline of the
light curve during the tail phase as illustrated in Figure
\ref{fig:stella_09dc_tail_highcsm_lc}.
However, the higher compression is achieved
by a stronger deceleration of the ejecta material, inducing an elemental
distribution which seems to be inconsistent with observations for \sndc{} (see
Figure \ref{fig:stella_09dc_tail_highcsm_elems}).
These findings put a strain on the otherwise quite successful interaction
model. The strong trapping during the tail phase seems to require an intense
compression of the ejecta, only achievable by a strong reverse shock. This
inevitably leads to a strong deceleration of the ejecta material, which seems
to be stronger than observed.

\section{Discussion}
\label{sec:discussion}

By performing a series of \stella calculations, we have identified a number of
generic consequences of fairly normal \snia{} exploding within a dense compact
carbon and oxygen rich envelope. Having presented these findings in the
previous sections, we briefly discuss a progenitor scenario which may provide a
CSM environment as investigated here. We also briefly re-examine our findings
in light of some limitations of our approach and finally assess a possible link
between interacting \snias{} and \sch{} explosions.

\subsection{CSM Origin}
\label{sec:csm_origin}

Within the context of \snias{}, the double degenerate mechanism is the most
plausible scenario to provide a CSM configuration and composition as
investigated in this work. In contrast to the recent advancement and success of
the violent merger scenario, involving a prompt explosion
\citep[e.g.][]{Pakmor2010, Pakmor2012, Moll2014}, we consider the classical
mechanism \citep{Webbink1984}, in which the secondary WD is completely
disrupted before a thermonuclear runaway can set in, as the most probable
candidate.  \cite{Yoon2007} showed that in this scenario, the subsequent
accretion of the disrupted secondary leads to the formation of a hot
nearly-spherical hydrostatic atmosphere and an extended thick accretion disk.
This material would qualitatively qualify for the carbon and oxygen rich CSM
investigated in this study.  However, we caution that it is still a matter of
ongoing debate whether such a `slow merger' would eventually lead to a
thermonuclear explosion and thus potentially to an \snia{} event or whether an
accretion induced collapse would be triggered \citep[see
e.g.][]{Yoon2007,Shen2012}. Nevertheless, this scenario has been investigated
previously, e.g.\ by \cite{Raskin2014}. Such explosions are often linked to
`tamped detonations', studied extensively by \citet{Khokhlov1993,Hoeflich1996},
and have been suggested as models for \sch{} \snias{}
\citep[e.g.][]{Scalzo2014,Raskin2014}.

\subsection{Geometry}
\label{sec:discussion_geometry}

All calculations presented here have been performed assuming perfect spherical
symmetry. This assumption naturally not only applies to the geometry of the
supernova ejecta but extends also to the CSM configuration. These
simplifications, which are owed to \stella's restriction to one-dimensional
spherically symmetric computational meshes, are acceptable for wind-like CSM
scenarios. However, as detailed above, the specific CSM composition considered
here suggests an association with the double-degenerate scenario, in particular
with the `slow-merger' model.

Even though the material in the immediate vicinity of the primary WD and thus
the explosion site is expected to establish an extended, nearly spherical
envelope, a prominent thick disk component should form as well
\citep{Yoon2007}. Consequently, strong deviations from spherical symmetry are
expected at least in this part of the circumstellar environment. Ejecta--CSM
interaction occurring in this configuration may lead to different signatures
along different lines of sight. The importance of varying column optical depth
along different lines of sight in the context of the most luminous \snias{} has
already been highlighted by \citet{Hillebrandt2007} and was seen, for example,
in the tamped detonation calculations by \citet{Raskin2014}. In the context of
\sch{} events, such line-of-sight effects may explain part of the diversity of
the class of objects. In addition, due to the anisotropic CSM distribution, the
reverse shock may have variable strength in different regions, leading to
various degrees of deceleration of the ejecta material.  Again, this may help
resolve the difficulties reported in the previous section concerning the tail
phase, namely achieving a sufficiently strong $\gamma$-trapping without
decelerating the ejecta material too drastically.  However, we emphasize that
the exact consequences of these geometrical effects are hard to predict. In
fact, they cannot be adequately captured with our simple one-dimensional
approach. Instead a fully-fledged multidimensional radiation hydrodynamical
treatment, capable of accounting for a multitude of line interactions would be
required. In this respect, Monte Carlo-based approaches, such as
\citet{Noebauer2012,Roth2015,Harries2015}, hold some promise, but the time
scales constraint discussed in Section \ref{sec:stella} may prove problematic.
Moreover, many of the cited methods have not yet been generalised to
multidimensional geometries or are designed for applications that are quite
different from interacting supernovae. In light of these considerations of
geometrical limitations, our study should be viewed as a first exploration,
investigating the overall effect of ejecta--CSM interaction and its general
plausibility as a model for \sch{} objects.

\subsection{Compactness}
\label{sec:compactness}

In the current investigation, the \stella calculations have been started at
$t_{\mathrm{exp}} = \SI{1e4}{s}$ after explosion for reasons of computational
convenience. Thus, before interacting with the CSM, the ejecta have already
expanded significantly leading to a substantial drop in density.  Within the
`slow-merger` picture, however, the ejecta--CSM interaction would be expected
to set in at earlier times, thus impliying a more compact configuration with
both the ejecta and the CSM at higher densities \citep[c.f.][]{Yoon2007}.

We briefly investigate the consequences of ejecta--CSM interaction occurring in
denser and more compact environments and explore specifically whether an
increase in the compactness may improve some of the current shortcomings of the
fiducial model (\modfid{}). First, the effects of reducing the time since
explosion are explored. This change is later combined with an increase in the
compactness of the CSM envelope. During the entire exploration, the total CSM
mass remains unchanged.

\begin{figure}
  \includegraphics{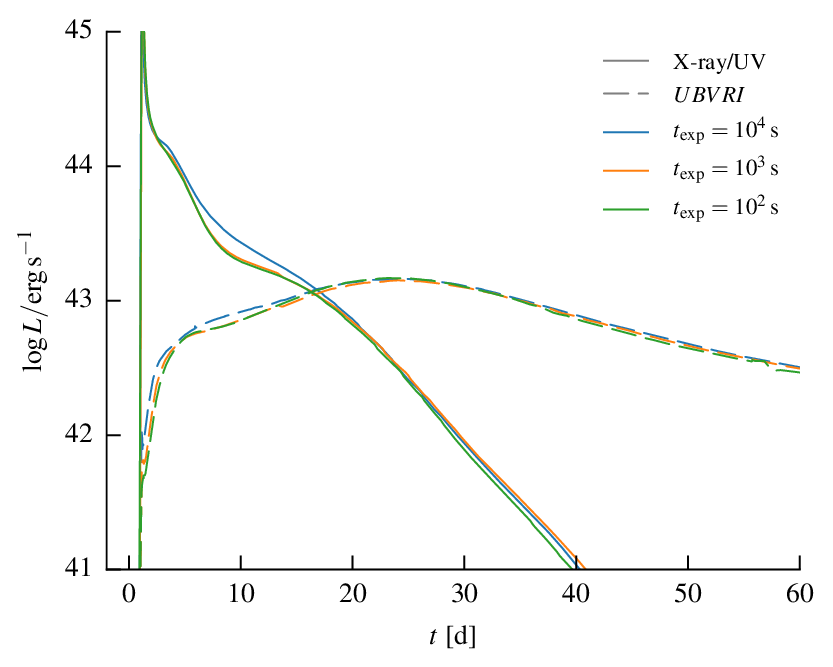}
  \caption{X-ray/UV and $UBVRI$ light curves for the calculations
    \modfid{}, \modfidearly{} and \modfidearlier{}. In this series, the time at
    which the \stella calculations are started is reduced from $10^4$ to $10^3$
    and $10^2\,\mathrm{s}$, resulting in increasingly more compact and dense
  ejecta. The extent of the CSM envelope remains unchanged.}
  \label{fig:lightcurve_texp_var}
\end{figure}

\begin{figure}
  \includegraphics{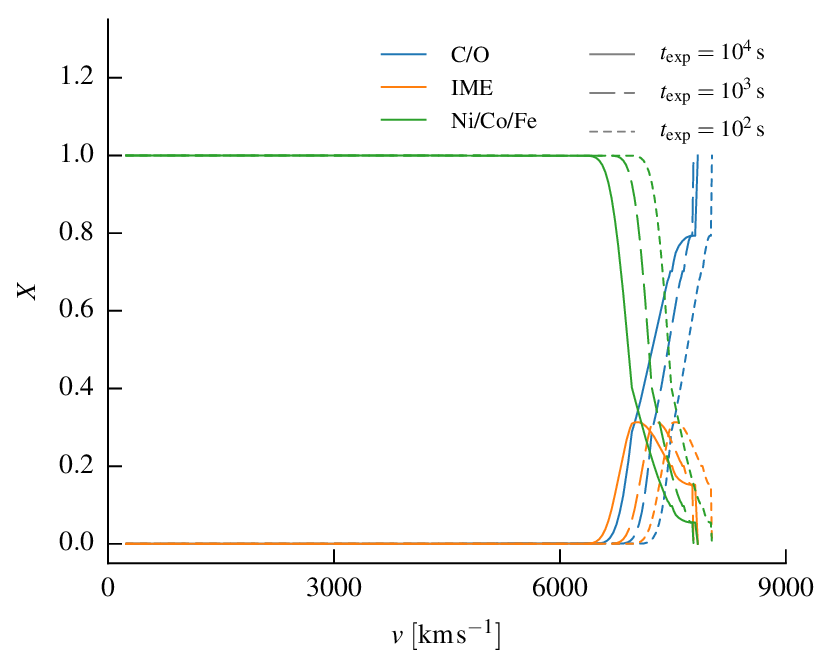}
  \caption{Elemental distribution at $t = \SI{10}{d}$ for the model series
  shown in Figure \ref{fig:lightcurve_texp_var}. The decelerating effect of the
reverse shock becomes slightly weaker as the time since explosion decreases and
thus the ejecta density increases.}
  \label{fig:texp_elements}
\end{figure}

Relying on the homologous expansion laws, the reduction of the time since
explosion leads to a density increase in the ejecta according to $\rho(t) =
\rho(t_0) (t_0/t)^3$.  Here, $t_0$ denotes an arbitrary reference time. This
boost of the ejecta density has only small effects on the amount of energy
generated in the ejecta--CSM interaction and only induces insignificant changes
in the X-ray/UV and $UBVRI$ light curves. Figure \ref{fig:lightcurve_texp_var}
illustrates this behaviour in terms of the emergent $UBVRI$ light curves for
the calculations \modfid{}, \modfidearly{}, \modfidearlier{} which were started
at $10^4$, $10^3$ and $\SI{.e2}{s}$ after explosion, respectively.
Due to the changing ratio of $\rho_{\mathrm{ejecta}}$ and $\rho_{\mathrm{CSM}}$,
the deceleration of the ejecta material induced by the reverse shock decreases
slightly in this series as shown in Figure \ref{fig:texp_elements}.

\begin{figure}
  \centering
  \includegraphics{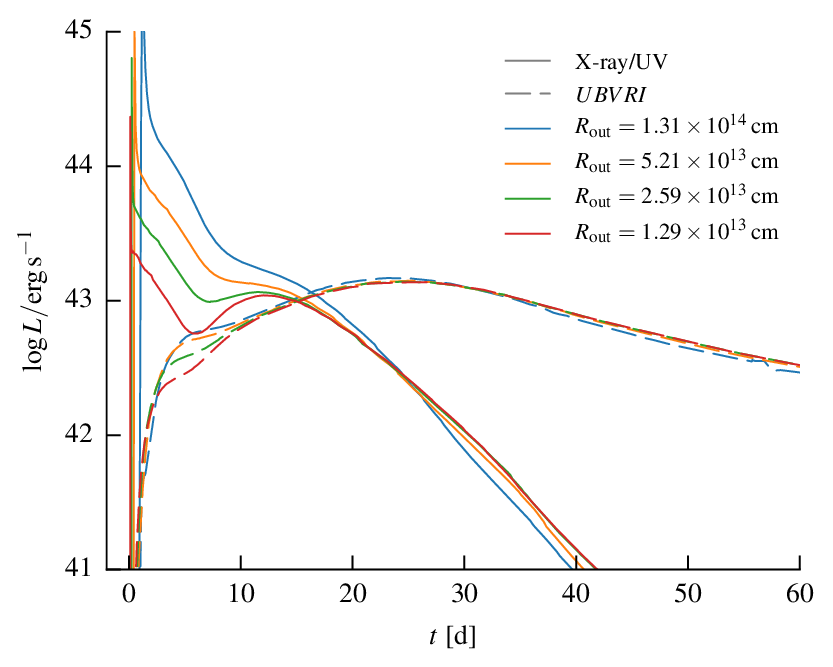}
  \caption{X-ray/UV and $UBVRI$ light curves for increasingly more compact
    ejecta-CSM configurations. In particular, the results of the calculations
    \modfidearlier{}, \modfidesmall{}, \modfidesmaller{} and \modfidesmallest{}
  are shown. The decreasing release of interaction radiation is a consequence
of the reduced size of the emitting system.}
  \label{fig:lightcurve_rhocsm_var}
\end{figure}

Based on the model \modfidearlier{} with $t_{\mathrm{exp}} =
\SI{.e2}{s}$ we complete the exploration of more compact configurations by
successively reducing the extent of the CSM envelope while keeping the enclosed
mass constant. In this process, the CSM density is boosted. Following the model
evolution with \stella reveals that still large amounts of energy are released
during the ejecta--CSM interaction. However, the interaction signature becomes
successively weaker, particularly in the X-ray/UV regime but also in the
$UBVRI$ light curve, as shown in Figure \ref{fig:lightcurve_rhocsm_var}.
During the early evolution, the transition between the optically thick and the
optically thin regime occurs deep inside the CSM envelope in the fiducial model
\modfid{}. By increasing the CSM density and reducing the outer CSM radius,
this transition region is pushed closer to the surface of the CSM envelope. As
a consequence, the radiation generated in the ejecta--CSM interaction breaks
out of the system at smaller radii and less total energy is overall emitted.

This exploration demonstrates that within the simplified one-dimensional
treatment and with the global parameters adopted here, a simple increase of the
compactness of the ejecta--CSM configuration does not lead to an increase in
the emitted radiation energy. However, this exploration should not be
interpreted as a detailed investigation of the plausibility of the
'slow-merger' model in the context of interacting SNe Ia, most notably for the
inadequate treatment of its complex geometrical properties (c.f.\ Section
\ref{sec:discussion_geometry}) and due to the limited exploration of the
involved parameters.

\subsection{CSM Composition and Line List}
\label{sec:discussion_linelist}

It has already been highlighted at the outset (see Section \ref{sec:stella}),
that \stella sacrifices some details in the radiative transfer treatment in
favour of a fully coupled radiation hydrodynamical solution strategy. With
respect to modelling \snias{}, the comparably small number of atomic line
transitions included in the opacity calculation is a significant limitation. In
a typical \stella calculation, as presented here, $\approx 1.5 \times 10^5$
line transitions are taken into account. In contrast, dedicated pure radiative
transfer simulations of \snia{} ejecta, for example performed with \artis
\citep{Kromer2009}, use up to millions of line transitions to accurately follow
radiative transfer and the frequency redistribution occurring within \snia{}
ejecta. 

As a consequence of relying on the small line list of \stella, an important
contribution to the overall opacity within the model may be missed. The omitted
processes may provide some of the necessary reprocessing of energetic radiation
produced in the ejecta--CSM interaction into the optical bands. The exploratory
test calculations performed with varying CSM compositions should be
viewed in this light. By enhancing the content of non-carbon-oxygen
material in the CSM, we boost the bound-bound opacity contribution and thus
very crudely approximate the effect of using a more complete line list. We note
that by using the prescription, we also enhance the bound-free contribution to
the total opacity and may influence the overall plasma state in the CSM.

\begin{figure}
  \centering
  \includegraphics{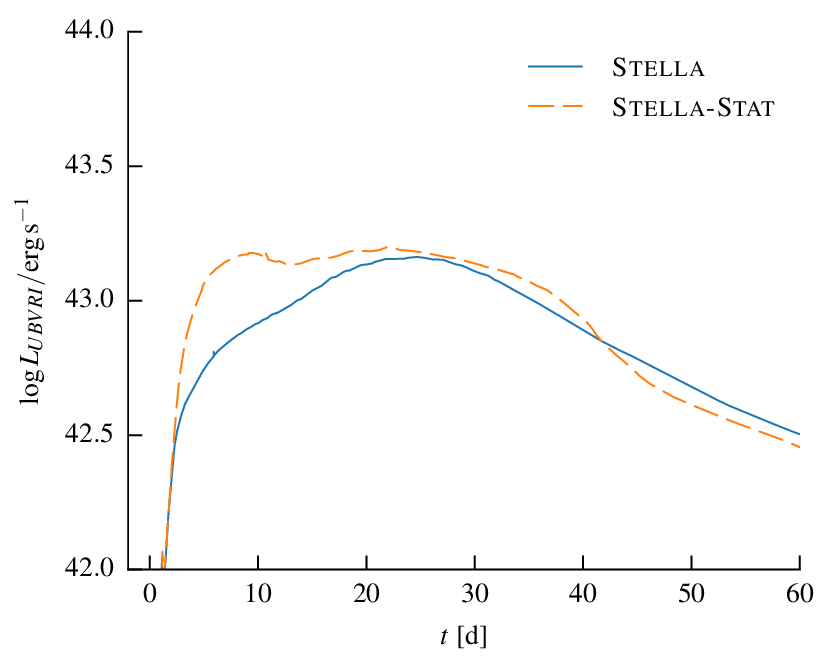}
  \caption{Comparison between the $UBVRI$ light curve of the interaction model
    \modfid{}, calculated with \stella (155000 lines) and \stellastat (26
    million lines). The result from the latter clearly shows the importance of
    the line opacity for reprocessing the energetic radiation into the optical
    bands and thus qualitatively supports our tests with enhanced CSM opacities
    presented in Section \ref{sec:peak}.} \label{fig:stellastat_lightcurve}
\end{figure}

Nevertheless, using high amounts of non-carbon-oxygen material in the
CSM qualitatively seems to be a valid representation of a larger line list.
This statement is supported by test calculations performed with an experimental
version of \stella \citep{Sorokina2016}. With this variant of the code
(referred to as \stellastat in this work), a large number of atomic line
transitions may be efficiently treated using a statistical approach.  Repeating
the simulation of the standard setup of the interaction model with a
$f_Z = 1$ CSM composition and 26 million lines, taken from
\citet{Kurucz1993}, shows that the optical luminosity stemming from the
ejecta--CSM interaction around peak is boosted compared to the results obtained
with the standard \stella version and 155000 lines (see Figure
\ref{fig:stellastat_lightcurve}).
However, we repeat that the \stellastat version is still under active
development and not yet fully tested on different SN types.

From the exploratory test presented in Section \ref{sec:peak} and the
preliminary calculations performed with \stellastat, we conclude that some of
the discrepancies we see in the fiducial calculation with a $f_Z=1$
CSM composition, \modfid{}, may actually be attributed to the use of an incomplete line
list and consequently to simplifications adopted in the radiative transfer
treatment of \stella.  However, the few calculations performed with
\stellastat, are not sufficient to confidently assess whether these effects
alone may fully account for the gap between model and \sndc{} observations
seen in Section \ref{sec:cmp_09dc}.

\subsection{\sch{} SNe from \snias{} within dense Carbon-Oxygen Envelopes?}
\label{sec:discussion_superch}

As the detailed comparison with \sndc{} demonstrated (see Section
\ref{sec:cmp_09dc}), the proposed interaction scenario is able to explain a
number of characteristic features of such \sch{} \snias{}. The decelerating
effect exerted onto the ejecta material by the reverse shock provides a natural
explanation for the low line velocities observed in some \sch{} objects. In
particular, the final elemental composition of the ejecta material observed in
the \stella simulations is compatible with the low line velocities measured for
\sndc{}. In addition, many properties of the optical light curve, such as a
long rise and a slow decline, can be attributed to the density enhancement
following the shock compression in the ejecta--CSM interaction. Specifically
with respect to \sndc{}, the light curve shape was nearly perfectly reproduced
in the \stella calculations.  Finally, the composition of the CSM and the fact
that it is quickly swept up by the ejecta, provides a natural explanation for
the absence of any tell-tale interaction signatures in the observational data,
such as narrow or intermediate-width hydrogen emission lines. Also, the high
carbon content in the CSM could explain the strong and persistent carbon lines
seen in many \sch{} SNe and facilitate dust formation which is suggested to
occur in a number of these objects.

Despite the many successes of the interaction scenario, achieving a significant
boost in the optical peak luminosity during the radiation hydrodynamical
calculations proves challenging. The explorations performed in the this work
suggest that the reprocessing efficiency of highly energetic radiation is key
for this. As demonstrated in Section \ref{sec:compactness}, however,
simply increasing the compactness of the ejecta--CSM configuration does not
help. Instead, the associated column optical depth associated with the
reprocessing efficiency may for example be increased by higher CSM opacities.
As detailed in Section \ref{sec:discussion_linelist}, shortcomings of the
radiative transfer treatment in \stella could be partly responsible for too low
a reprocessing efficiency in the current calculations.  Also, line-of-sight
effects, which are expected to occur when realistic geometries are considered
(c.f.\ Section \ref{sec:discussion_geometry}), may help in achieving a boost.
Irrespective of these considerations, a number of \sch{} objects are less
luminous than \sndc{}, for example SN~2006gy or SN~2012dh.  Their optical
light curve peaks may already be reproduced by models presented here,
independently of the consequences of omitted radiative transfer effects and
geometrical influences.

In summary, given the many successes of the current interaction scenario with
respect to reproducing properties of \sndc{}, we consider it a promising
candidate for explaining \sch{} \snias{}. Additional explorations, which sample
a wider parameter range, undertake detailed spectral synthesis calculations and
address the inherent multidimensional geometry, are recommended to further
support this claim.

\section{Summary and Conclusions}

In this work, we have investigated the consequences of \snias{} being
surrounded by a thick carbon-oxygen rich CSM envelope. Our main interest rested
on studying the effect of the ensuing ejecta--CSM interaction on the ejecta
evolution and on the emergent light curve. Here, we focussed on a particular
realisation of this scenario, which has previously
\citep[c.f.][]{Taubenberger2013} been identified as a promising candidate for
the explanation of \sch{} \snias{}, in particular for \sndc{}. 

We followed the evolution of the considered model using the radiation
hydrodynamical code \stella \citep{Blinnikov1993,Blinnikov1998,Blinnikov2006}.
At the ejecta--CSM interface, a strong forward and reverse shock form, sweeping
up the CSM and compressing the ejecta material, respectively.  Due to the shock
heating processes, a substantial amount of energy is injected into the
radiation field. However, this occurs mostly in the X-ray and UV regime and
only part of it is shifted into the optical $UBVRI$ bands due to reprocessing
in the CSM. As a consequence, no significant boost in the maximum $UBVRI$
luminosity is observed compared to a corresponding `bare' \snia{} explosion.
However, we find that the ejecta--CSM interaction leads to a longer rise and a
significant broadening of the light curve. Past maximum, a slower decline of
the optical light curve is observed due to the increased $\gamma$-trapping
induced by the compression of the ejecta material by the reverse shock. Apart
from the density increase, the ejecta material has also been significantly
decelerated, leading to an extended velocity plateau in mass space.

Despite the missing luminosity boost in the fiducial model, we find that an
increase in the light output in the optical regime may still be achieved if the
reprocessing efficiency in the CSM is enhanced. Not observing this in the
current calculations may simply be a consequence of simplifications adopted in
the radiative transfer scheme implemented in the standard version of \stella.
In particular, using a realistic near-complete line list significantly
increases the opacity and the reprocessing efficiency. We performed simple test
calculations with an enhanced CSM opacity and by carrying out
calculations with an experimental \stella version, capable of treating millions
of atomic line transitions, supporting these statements. During the light curve
decline phase, the increased $\gamma$-trapping is a direct consequence of the
compression due to the reverse shock. Thus, its strength regulates the optical
luminosity during these phases and an additional boost may consequently be
achieved here, if the ejecta is, for example, embedded into a more massive CSM
envelope.

In addition to these generic findings, the interaction scenario successfully
reproduces a number of characteristic features of \sch{} \snias{},
despite the difficulty to produce a significant luminosity boost in the
optical light curve. When comparing to the prototype of this class, \sndc{}, we find that
the shape of the observed optical $UBVRI$ light curve is almost perfectly
matched. In particular, the long rise time, the broad maximum and the delayed
decline due to increased $\gamma$-trapping is well reproduced.  Moreover, the
deceleration of the ejecta material due to the reverse shock provides a natural
explanation for the low line velocities observed in \sndc{}.  The velocity
distribution of the different elements found in the model is compatible with
these observations. However, to establish a directly link, detailed spectral
synthesis would have to be performed, a task which cannot be easily carried out
with \stella. Despite these successes, the defining characteristic feature of
\sndc{}, namely the high light output is not fully reproduced in the numerical
calculations. However, as pointed out already above, limitations of the
radiative transfer treatment in \stella may be partly responsible for that.

Next to the compromises in the radiative transfer treatment, the restriction to
spherically symmetric configurations is a shortcoming of the current study. In
particular, if double degenerate scenarios are considered as the possible
origin of the CSM material, deviations form spherical symmetry may play an
important role. The exact consequences of these are difficult to predict, but
line-of-sight effects seem plausible. To some extent, these could provide a
natural explanation for some of the diversity seen in \sch{} objects.

In summary, despite some shortcomings of our current study, the initial
investigation of \snias{} interacting with a dense carbon-oxygen rich envelope
shows that this scenario may provide a viable explanation for \sch{} objects.
To further investigate this possibility and to better quantify the generic
consequences of ejecta--CSM interaction, we aim to continue our study of this
scenario in the future, specifically focussing on three points. Firstly, a
better sampling of the possible parameter space, both in terms of the ejecta
and the CSM properties, should be performed and thus different configurations
of the interaction scenario investigated. In addition, we aim to map results
from \stella simulations into dedicated radiative transfer approaches, such as
\artis \citep{Kromer2009} and \tardis \citep{Kerzendorf2014}, once homology is
approximately retained. This way, detailed colour light curves may be
determined and the spectral appearance of the interacting models predicted. In
parallel, it should be explored how alternative radiation hydrodynamical
methods, such as \mcrh \citep{Noebauer2012,Noebauer2015}, may be applied to the
interacting \snias{} problem, with the goal of eventually overcoming the
current limitation to one-dimensional geometries. These steps will help
establishing whether super-luminous \snias{} are interaction-powered and
whether and to which extent ejecta--CSM interaction is relevant for other
\snias{}.

\section*{Acknowledgements}

The authors would like to thank M.~Kromer, R.~Pakmor and S.~Sim for stimulating
and fruitful discussions. The authors also thank the anonymous reviewer for
valuable comments. This work has been supported by the Transregional
Collaborative Research Center TRR33 `The Dark Universe' of the Deutsche
Forschungsgemeinschaft and by the Cluster of Excellence `Origin and Structure
of the Universe' at Munich Technical University. The work of SB (development of
the code \stella) has been supported by a grant of the Russian Science
Foundation 14-12-00203, the work of ES (statistical approach for opacity
calculation) has been supported by a grant of the Russian Science Foundation
16-12-10519.







\end{document}